\documentclass[aps,prc,preprint,groupedaddress,preprint,floatfix]{revtex4}
\usepackage{graphicx}
\usepackage{bm}
\usepackage{amssymb}
\usepackage{epsf}
\usepackage{epsfig}

\def\gtorder{\mathrel{\raise.3ex\hbox{$>$}\mkern-14mu
\lower0.6ex\hbox{$\sim$}}}
\def\ltorder{\mathrel{\raise.3ex\hbox{$<$}\mkern-14mu
\lower0.6ex\hbox{$\sim$}}}

\begin{document}

%\linespread{0.8}

\title{	Measurements of the EMC effect in few-body nuclei at large $x$ }

\author{A. Daniel }
\affiliation{Dept. of Physics and Astronomy, Ohio University, Athens, OH }
\author{J. Arrington}
\affiliation{Argonne National Laboratory, Argonne, IL}
\author{D. Gaskell}
\affiliation{Jefferson Laboratory, Newport News, VA}

%\author{D. Gaskell}{  address={Jefferson Laboratory, Newport News, VA}, altaddress=(for the JLab E03$-$103 collaboration).}
% (for the JLab E03$-$103 collaboration)

\begin{abstract}

  Recent data on the EMC effect from  Jefferson Lab  experiment  E03103
  suggest that the nuclear dependence of the high $x$  quark  distribution
  may depend on the  local nuclear environment, rather than  being a purely
  bulk effect. We briefly discuss the analysis and results of the
  E03103 experiment focusing on the light nuclei and high $x$ region. In addition, we also  discuss a future experiment at the upgraded 12 GeV Jefferson Lab facility which will further investigate the role of the local nuclear environment and the influence of detailed nuclear structure on the modification of quark distributions.

\end{abstract}
\keywords      {EMC effect, structure functions, light nuclei}

\pacs{13.60.Hb,25.30.Fj,24.85.+p}

\maketitle

%%%%%%%%%%%%%%%%%%%%%%%%%%%%%%%%%%%%%%%%%%%%
%% MAINMATTER
%%%%%%%%%%%%%%%%%%%%%%%%%%%%%%%%%%%%%%%%%%%%

\section{Introduction}
The investigation of deep inelastic scattering (DIS) of leptons off the
nucleon is one of the most effective ways for obtaining fundamental information 
on the quark-gluon substructure of the nucleon. On the other hand, measuring nuclear structure
functions enables  us to probe the distribution of nucleons in a nucleus, as
well as the nuclear modification to the structure of a nucleon. Existing data
on nuclear DIS  shows that the quark  distributions in heavy nuclei are not simply the sum of the quark
 distributions  of the constituent protons and neutrons, as one might expect for
 a weakly bound system. This modification of the quark distributions in nuclei
 is known as the EMC effect.

Even though it has been more than two decades since the discovery of the EMC
effect ~\cite{originalemc} and there are extensive data on the $x$ and $A$
(mostly $A$ $\gtorder$ 12) dependence of the EMC
effect, there is no clear consensus as to its origin.  Existing data show a
universal shape for the EMC ratios in the valence region, and a weak
dependence on $A$. However, the nuclear dependence can be well fit with either a
simple $A$-dependent or density dependent form~\cite{Gomez:1993ri}, thus making
it difficult to test models of the EMC effect based on the $A$ dependence.  In
addition, the EMC effect at very high $x$ values is not well measured
because of the constraint that the data be taken in the DIS region.

Because the nuclear dependence appears to have contributions from multiple
sources, it has been difficult to determine exactly what causes the observed
behavior. The formalisms used to explain the observed effect range from
traditional nuclear physics descriptions in terms of pion exchange  or binding energy
shifts, to QCD inspired models such as dynamical rescaling, multi quark
clustering, de-confinement in nuclei, etc. (see, for example, Refs~\cite{Geesaman:1995yd,
  Norton:2003cb,Cloet:2006bq,Smith:2003hu}). Many models attempt to explain the observed
effect using the convolution formalism to incorporate nuclear physics effects. This is challenging due to the difficulty in obtaining detailed nuclear wave functions and one must often resort to mean-field calculations.
Another approach tries to explain the EMC effect using the changes to the
internal structure of the nucleon. As mentioned earlier, despite a quarter
century of experimental effort and much theoretical progress, no unique and
universally accepted explanation of the EMC effect has emerged and the
specific origins of the nuclear dependence remain a topic of debate. It should be noted that the conventional nuclear physics effects are not only significant at large $x$, but lead to a modification of the distributions at all values of $x$.  These effects are the dominant contributions at large $x$, but the limited data at
large $x$, coupled with the lack of precise data on light, easily calculable
nuclei, has made it difficult to determine how well these `traditional'
nuclear physics effects are being included in models of the nuclear structure
function.

\section{Results from Jefferson Lab experiment E03103}
\begin{figure}[htbp]
\includegraphics[height=80mm]{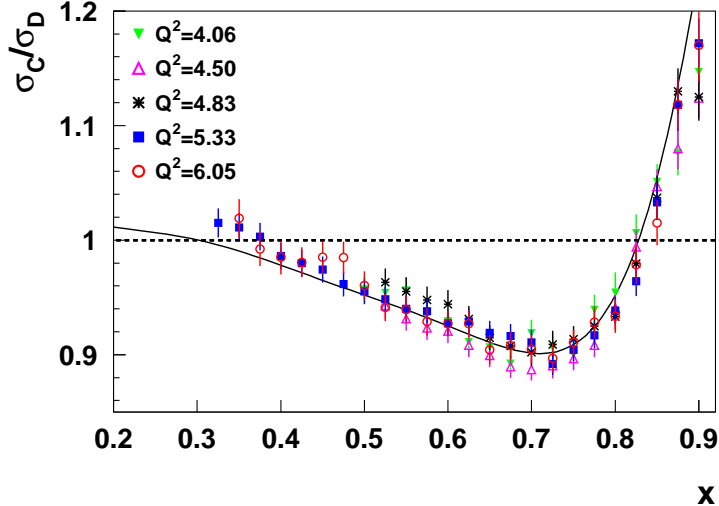}
\caption{ Shown are the  EMC ratios for the five highest Q$^2$ settings ( Q$^2$ quoted at $x$=0.75) for $^{12}$C. The solid curve is the SLAC E139 fit to the Carbon EMC ratio.\label{emc_c_q2dep_fig}}
\end{figure}

Experiment E03103~\cite{emcprop}, which ran in Hall C at Jefferson Lab, measured inclusive electron scattering cross sections from few-body and heavy nuclei. The primary goal was to make improved measurements of the EMC effect, focusing on large $x$, where Fermi motion and binding are believed  to be the dominant effects, and for light nuclei, where the uncertainties in the nuclear structure are smaller, thus reducing uncertainties in comparisons to calculations of the EMC effect. The majority of the data were taken with an electron beam energy of 5.766 GeV with beam currents ranging from 30 to 80 $\mu$A. The cryogenic targets $^{1}\mathrm{H}$, $^{2}\mathrm{H}$,$^{3}\mathrm{He}$,$^{4}\mathrm{He}$ and solid targets Be, C, Cu and Au were studied.  Data for all targets were taken at electron scattering angles of 40$^\circ$ and 50$^\circ$, covering a $Q^{2}$ range 3 (GeV/c)$^2$ at $x= 0.3$ and 6 (GeV/c)$^2$ at $x=0.9$. At high $x$, the kinematics were not in the conventional DIS region, so additional data at 4 other angles covering 18$^\circ $ to 32$^\circ $ were collected at a beam energy of 5.766 GeV.  Data were also collected for a detailed $Q^{2}$ dependency study at 4 angles on C and $^{2}\mathrm{H}$, at a beam energy of 5.01 GeV. Scattered electrons were detected in the High Momentum Spectrometer. Details about the experiment and the EMC ratio extraction can be found in  \cite{e03103prl,aji_thesis}.
\begin{figure}[htbp]
\includegraphics[height=57mm]{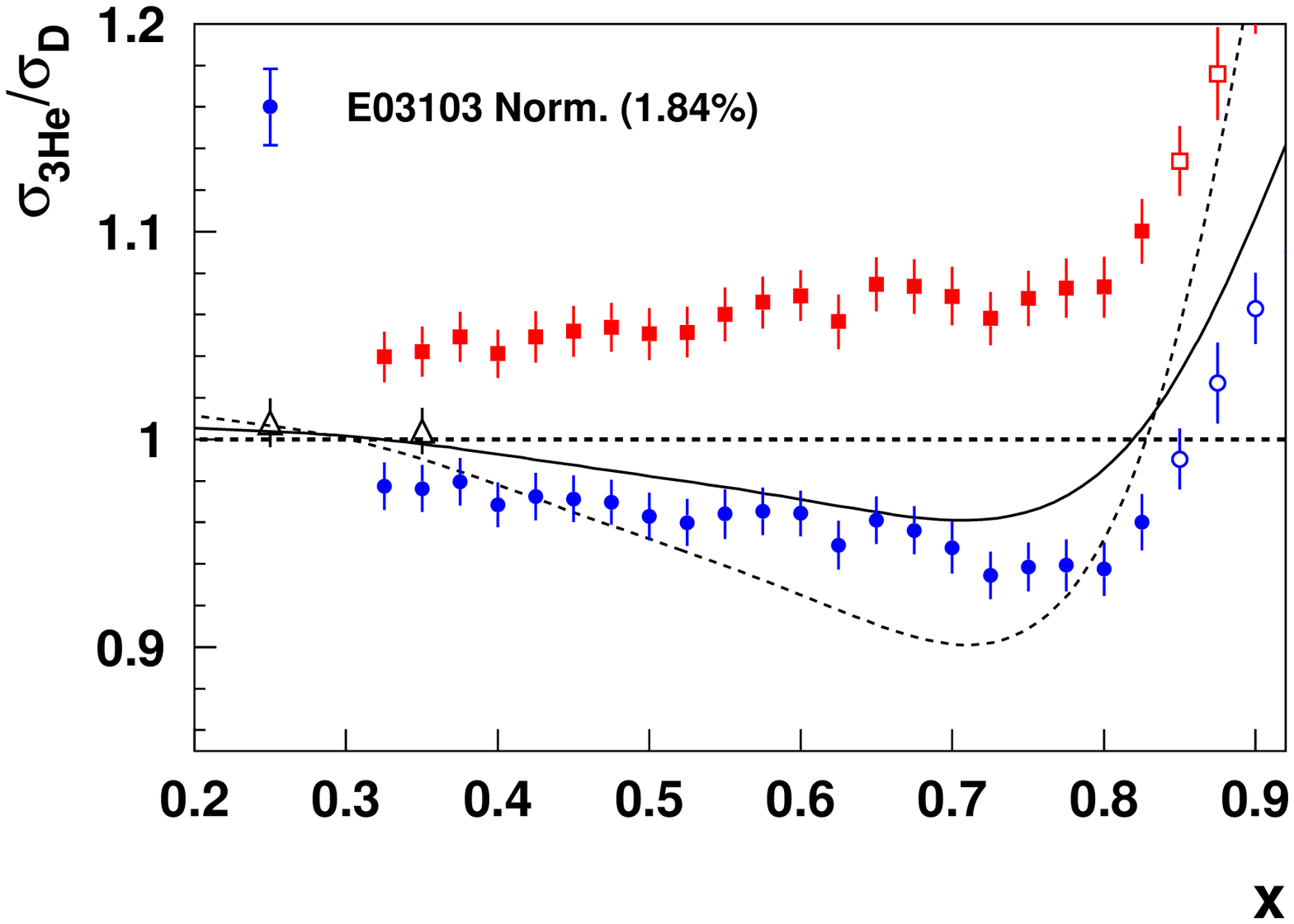}
\includegraphics[height=55mm]{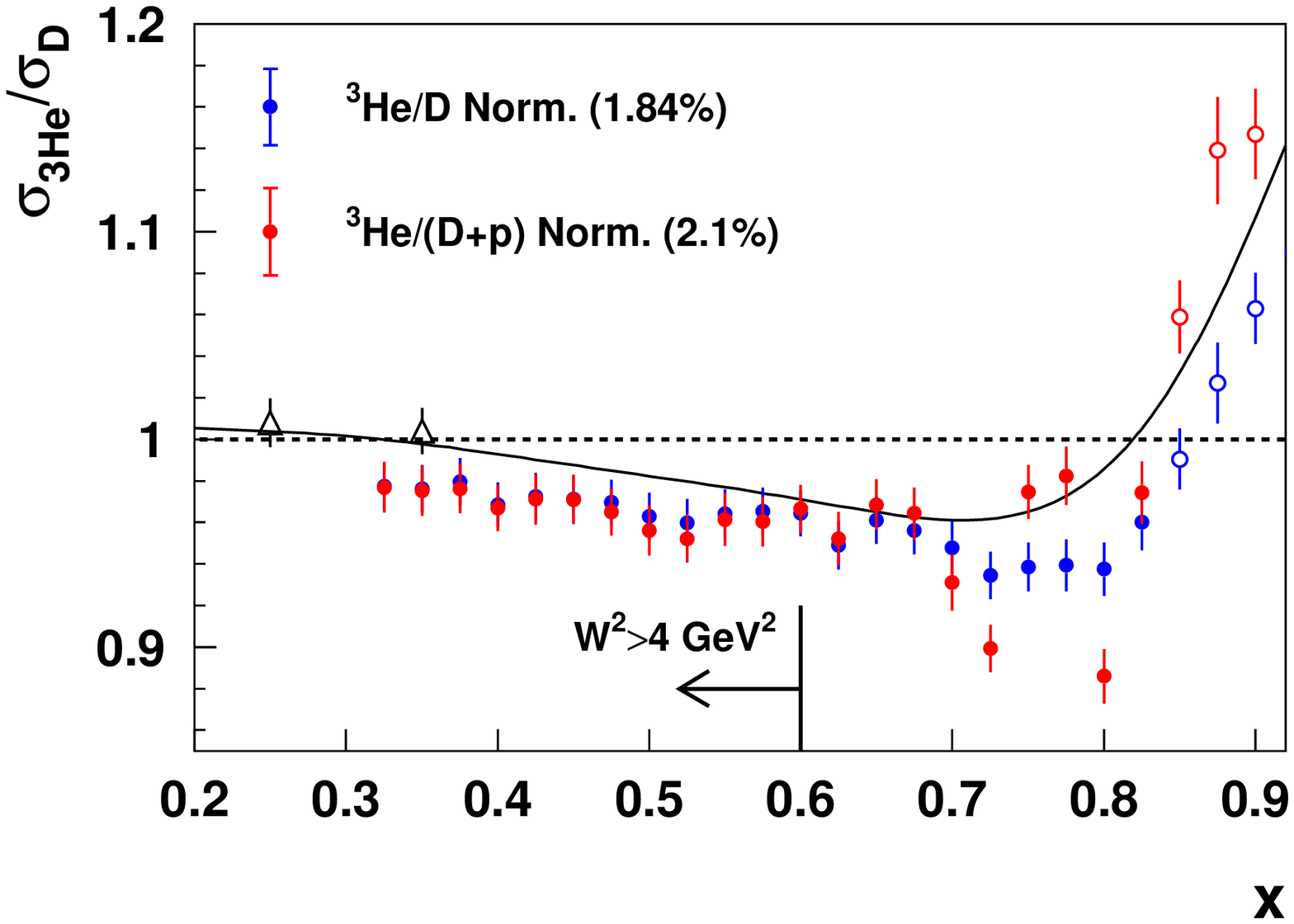}
\linespread{0.9}
\caption{The left panel shows the raw and isoscalar-corrected $^3$He/$^2$H
ratios, compared to the SLAC E139 fit for $^3$He (solid) and Carbon (dashed). The upper squares are the raw $^3$He/$^2$H ratios, while the bottom circles show the isoscalar corrected EMC ratio.  The triangles are the HERMES results~\cite{Ackerstaff:1999ac} which use a different isoscalar correction.  The right panel shows the same isoscalar $^3$He/$^2$H ratios (blue points) and the $^3$He/($^2$H+$^1$H) ratio (red points) extracted from E03103 data. The black curve is the  SLAC mass number dependent fit to $^3$He. Hollow symbols indicate data at $W^2 \leq 2$~GeV$^2$.}\label{he3_dprat.fig}
\end{figure}

The E03103 data are in the conventional DIS ($W^2 > 4$~GeV$^2$) region up to $x\approx 0.6$ as shown in Figure~\ref{kinem.fig}. While the JLab data were at somewhat lower $Q^2$ values than SLAC E139 (which took data mainly at 5 and 10~GeV$^2$), we obtained high precision at larger $x$ values by making measurements at somewhat lower $W^2$ values. Data were also taken at several energy and angle settings, to map out the $Q^2$ dependence of the ratios and cross sections in detail, allowing us to verify that our data were interpretable in the context of quark distributions in nuclei. Although there were indications \cite{Arrington:2003nt}  that the nuclear structure functions in the resonance region show the same global behavior as in the DIS region, this was explicitly tested using E03103 data. These studies show that the cross section ratios scale to low values of $Q^2$ and $W$. As seen in Figure~\ref{emc_c_q2dep_fig}, the C to  $^{2}\mathrm{H}$ ratio for the five highest $Q^2$ settings show no systematic $Q^2$ dependence in the cross section ratios, even at the largest $x$ values. This implies that the contribution from higher twist effects and additional scaling violation mechanisms to the ratios are small even in the high $x$ region.

E03103 provided the first measurement of the EMC effect for $^3$He in the valence region, as well as significantly improved data on $^4$He. Figure~\ref{he3_dprat.fig} shows the EMC ratio for $^3$He, with the lower-$x$ data from HERMES \cite{Ackerstaff:1999ac}. Since  $^3$He contains two protons and a neutron, the raw cross section ratios must be corrected for the proton excess to obtain the isoscalar corrected EMC ratios. In the figure, the upper squares are the raw $^3$He/$^2$H ratios while the bottom circles show the isoscalar corrected EMC ratios. There is a significant uncertainty in the neutron cross section in the large $x$ region, hence, the extracted EMC ratios are very sensitive to these isoscalar corrections. Previous measurements (SLAC E139)  used  a correction based on the high $Q^{2}$  measurements of the free $F_{2n}/F_{2p}$ ratio. Since  one is correcting the nuclear cross sections, one should be using the contributions of $F_{2p}$ and $F_{2n}$ to the nuclear structure function instead of using the free proton and neutron structure functions. E03103 results are extracted using global fits of free proton and neutron cross sections and broadened using  the $F_{2n}/F_{2p}$ ratio and the convolution procedure mentioned in \cite{Arrington:2008zh}. These distributions are then used to get the bound $F_{2n}/F_{2p}$ ratios in nuclei. It should be noted that the isoscalar corrections depends on $x$ and $Q^{2}$, and the $Q^{2}$ dependence is not negligible at large $x$ values.

In the case of $^3$He, one can avoid the uncertainty associated with the isoscalar correction, and thus better evaluate models of the EMC effect, by taking the ratio of $^3$He to ($^2$H+$^1$H) which allows comparisons to calculations that are independent of the neutron structure function. This is shown in the right panel of Figure~\ref{he3_dprat.fig}. Here,  the blue points show the isoscalar corrected  $^3$He/$^2$H  ratios while the red points shows the $^3$He/($^2$H+$^1$H) ratio  extracted from E03103 data. As one can see both  ratios have good agreement below $x \approx 0.65$, however, the resonance structure in the proton is not washed out, and so the extended scaling observed in nuclei is not as effective, limiting the useful range for this ratio to $x \ltorder 0.65$.
\begin{figure}[htbp]
\includegraphics[height=75mm]{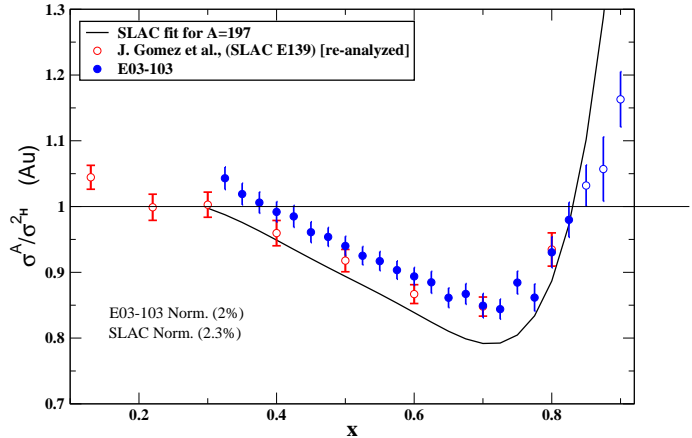}
\caption{Preliminary results for $^{197}$Au EMC ratios. The blue filled circles  represents  E03103 results while the red empty circles represent SLAC E139 results. The black curve is the  SLAC $A$ dependent fit to $^{197}$Au.\label{emc_au_x_fig}}
\end{figure}

One additional correction that needs to be applied to the data, especially for heavy nuclei, is due  to the acceleration of the incoming electrons and deceleration of outgoing electrons in the Coulomb field of the target nucleus (so called Coulomb corrections). These corrections are typically ignored because the magnitude of the correction factor is small at very high energies. We estimated these correction factors using an improved version of the Effective Momentum Approximation (EMA) as described in \cite{aste_ccor}. This was found to be very small for the light nuclei, however, it was also observed that for heavy nuclei, the impact of Coulomb distortion was not negligible, even for the SLAC measurements~\cite{Solvignon:2009it}. For the JLab data this correction is found to be about 18\% for $^{197}$Au at high $x$ near the quasi-elastic peak~\cite{aji_thesis}.  It should be noted that, like the isoscalar corrections, Coulomb corrections are also $Q^{2}$, $x$ and $A$ dependent. Because of their $A$ dependence, proper treatment of these corrections are important in extrapolations to nuclear matter or any global analysis of the data regarding the EMC effect. Preliminary results for the $^{197}$Au EMC ratios are shown in Figure~\ref{emc_au_x_fig} along with the data from SLAC E139 which has been reanalyzed to include Coulomb distortion effects.

\begin{figure}[htbp]
\includegraphics[height=75mm]{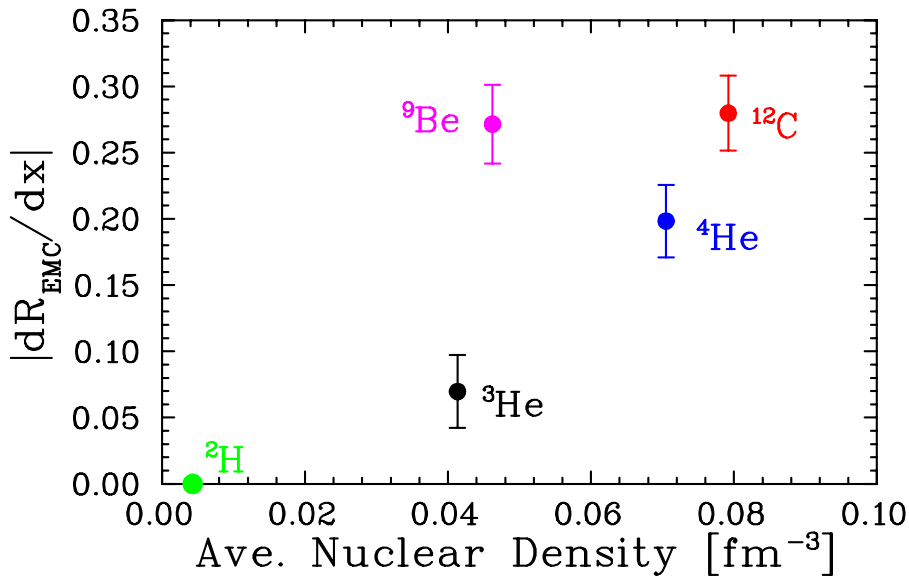}
\caption{The figure shows the slope of the isoscalar EMC ratios for 0.35 $< x < $0.7 as a function of scaled nuclear density.}\label{size_emc.fig}
\end{figure}

The full results for the EMC ratios for $^3$He, $^4$He, Be and C are available in Ref.~\cite{emcprop,e03103prl}. The results for all these nuclei are consistent with the SLAC measurements, with much better precision for $^4$He and new measurements for $^3$He in the valence region. Figure~\ref{size_emc.fig} shows the size of the EMC effect for these light nuclei as a function of the scaled nuclear density.  To avoid contributions from the normalization uncertainty in the measurements, especially important for nuclei with a small EMC effect, we quantify the size of the EMC effect based on the slope of the EMC ratio in the linear region between $x=0.35$ and $x=0.7$.  For the nuclear density of these light nuclei, the assumption of a uniform sphere density distribution, as used in the SLAC fits, is not a good approximation.  We take the density distributions calculated from the \textit{ab initio} Green's Function Monte Carlo calculations~\cite{Pieper:2001mp} to calculate the average density for each nucleus.  We choose to scale down this average nuclear density by a factor of ($A$-1)/$A$, based on the idea that each  nucleon is only influenced by the other ($A$-1) nucleons.
\\
\\
Note that $^3$He and $^9$Be have similar average densities, which are significantly lower than the densities for $^4$He and $^{12}$C.  If the data behaved according to the $A$-dependent fit~\cite{Gomez:1993ri}, one would expect $^3$He and $^4$He to be similar in magnitude and roughly a factor of two lower than $^9$Be and $^{12}$C.  The density-dependent fit would predict similar values for $^4$He and $^{12}$C, with significantly lower effects for $^3$He and $^9$Be.  The E03103 results show $^3$He has a much smaller EMC effect than either $^4$He or $^9$Be, suggesting that neither the mass-dependent or density-dependent parameterizations describe these light nuclei.

While neither of these simple models for the scaling of the EMC effect accurately describe the light nuclei results, the data are consistent with the idea that the nearby nucleons are most important.  The structure of $^9$Be includes a significant component with two alpha clusters and one excess neutron.  In this picture, most of the nucleons  are bound in these tight clusters, and thus the local environment of the nucleons is similar to $^4$He, even though the average density is quite low.  This suggests that these clustering effects and the local environment may be important, and clearly shows that calculations of the EMC effect must take into account the detailed nuclear structure. There has been some recent work by Hirai and collaborators~\cite{Hirai:2010xs} aimed at investigating the possible connection between the DIS structure functions to the clustering structure in nuclei. 

\section{Future measurements at Jefferson Lab (experiment E10-008)}
The results from E03103 emphasize the need for precise measurements on a range of well understood nuclei. The nuclei included in E03103 were sufficient to examine the simple $A$-dependent and density-dependent scaling models, and show that detailed calculations including realistic nuclear structure will be important in explaining the EMC effect. A recently approved experiment \cite{e1210008_proposal} will provide further information on the detailed behavior of the observed nuclear dependence with an expanded set of light nuclei, including nuclei with significant clustering contributions. 
\begin{figure}[htbp]
\includegraphics[height=85mm]{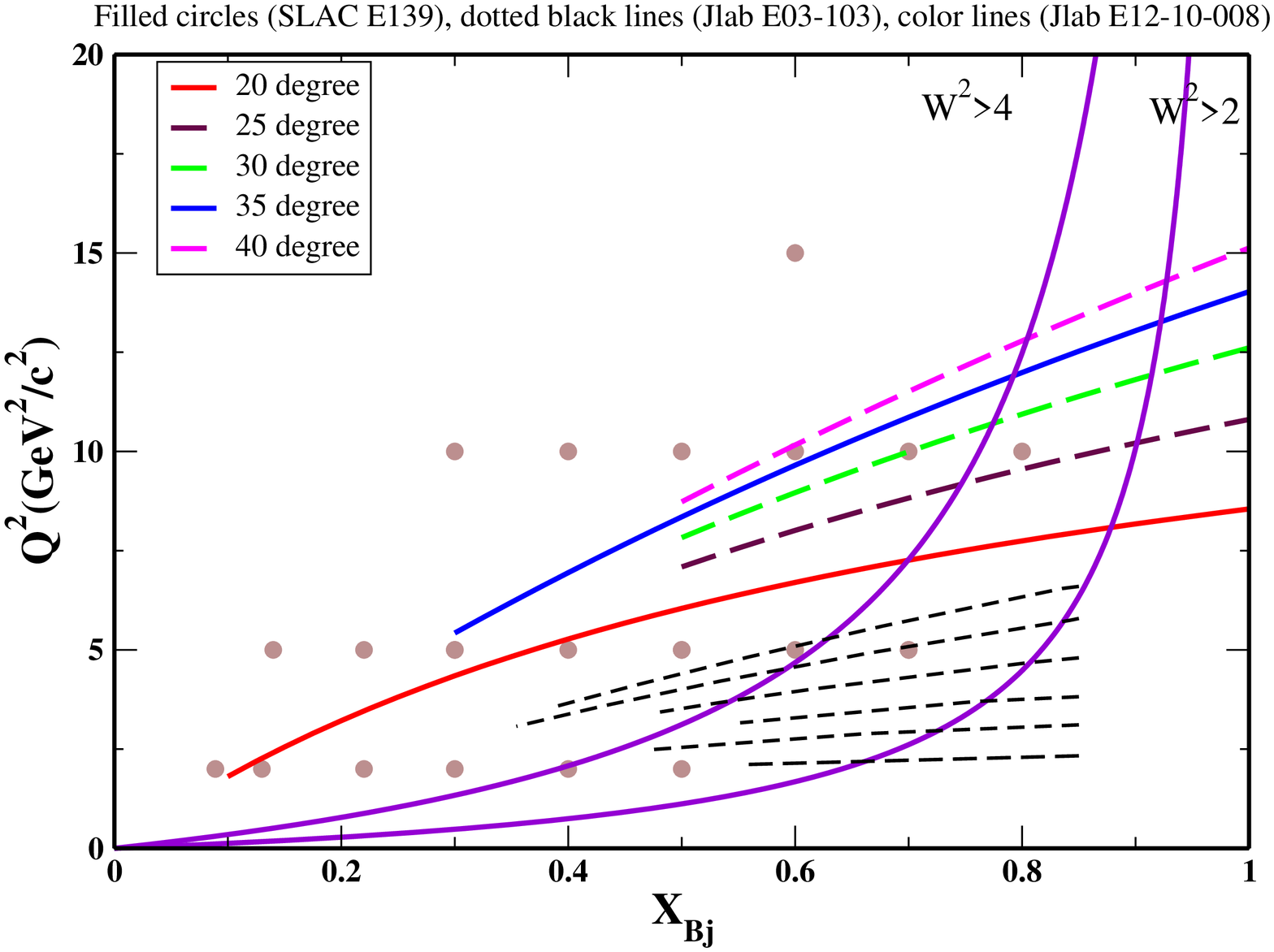}
\linespread{0.9}
\caption{Kinematics of the proposed measurements at 12 GeV. The black dotted lines are the kinematics from  E03103, and the solid red and blue lines are the primary kinematics for the E10-008 experiment. Data on a small subset of targets at additional angles will be used to study the $Q^2$ dependence of the ratios (shown as dashed lines). The solid purple lines correspond to contours of fixed $W^2$ values as indicated in the plot. SLAC E139 kinematics are shown as filled circles.}\label{kinem.fig}
\end{figure}

This new experiment will provide data on hydrogen, deuterium, $^3$He, $^4$He, $^{6,7}$Li, $^9$Be, $^{10,11}$B, $^{12}$C,$^{40,48}$Ca, and  $^{63}$Cu. While $^4$He and $^9$Be are especially well suited for separating the $A$-dependent and density-dependent pictures, scaling based on the \textit{local} density,  as estimated from two-body correlation functions from the GFMC calculations, yield predictions in between the $A$ and $\rho$ dependence. For $^{6,7}$Li, the local density picture predicts much smaller  EMC effect compared to other models.

The kinematics of  the  proposed measurements are shown in Fig.~\ref{kinem.fig}. The higher beam energy available after the upgrade of Jefferson Lab will increase the region of precise scaling to larger $x$ values; going from $x$=0.6 to 0.8 for $W^2>4$~GeV$^2$, and up to $x=0.92$ for $W^2>2$~GeV$^2$, where precise scaling was observed at the lower $Q^2$ values of E03103. Extending the measurements down to $x \approx 0.1$ will let us better compare the shape ($x$ dependence) of the EMC effect in these nuclei.

In addition, measurements at higher beam energy will reduce the uncertainty related to the isoscalar correction for $^3$He and allow  for better comparisons of the $x$ dependence of the EMC effect in light nuclei. As shown in Figure~\ref{he3_dprat.fig}, one can compare the  isoscalar corrected $^3$He/$^2$H ratios to $^3$He/($^2$H+$^1$H) ratios. For the proposed measurements, the resonance structure shifts to larger $x$ values and becomes much smaller, yielding a negligible deviation from the DIS limit up to $x$=0.85. Thus, the data on $^3$He  can be precisely compared to detailed calculations, without the uncertainty associated with knowledge of the neutron structure function. Note that, this can also be done for other light, non-isoscalar nuclei (e.g. $^7$Li), to allow for calculations that are insensitive to knowledge of the neutron structure function.

Including additional non-isoscalar targets will allow for additional tests of the EMC effect. There have been recent suggestions of a significant quark flavor dependence in the EMC effect~\cite{Cloet:2006bq}. The isospin dependence of the interaction leads to different degree of modification for up and down quark distributions. E10-008 will have  the ability to look for a possible flavor dependence of the EMC effect by measuring the EMC effect in the calcium isotopes. Measurements of $^{40}$Ca and $^{48}$Ca will provide a significant variation of the n/p ratio in the nucleus, while maintaining a comparison between nuclei of similar mass and density. Also, comparisons of nuclei which differ by just one nucleon (for example $^{11}$B-$^{10}$B, $^{7}$Li-$^{6}$Li,  $^{12}$C-$^{11}$B) will allow the extraction of the structure function of a single nucleon in the nucleus. This can be used to  check the isoscalar corrections applied in these nuclei.

\section{summary}
We have measured the nuclear dependence of the EMC effect for a series of light nuclei, and the  data from E03103 have provided important new information on the EMC effect. We have improved the precision of the data at high $x$ by relying on the extended scaling that we observe from our data. Also, this experiment provides the first valence-region measurement of EMC effect in $^3$He. E03103 provides benchmark data for models of the EMC effect, allowing comparison to few-body nuclei with well understood structure, as well as providing high precision data at larger $x$ values than were available from the SLAC measurements. Experiment E10-008 will further investigate the influence of nucleon's local environment on the observed nuclear effects. These data will improve on existing JLab measurements of the EMC effect by extending precise extraction of the EMC ratios to larger
$x$ values, and by making additional measurements on light nuclei to provide better data for constraining calculations of nuclear effects in these well understood nuclei.

\section{Acknowledgments}
 This work is supported in part by research grant PHY-0653454 from the National Science Foundation and in part by the U.S. Department of Energy grant DE-AC02-06CH11357. The Southeastern Universities Research Association operates the Thomas Jefferson National Accelerator Facility under the U.S. Department of Energy  contract DE-AC05-84ER40150.

%%%%%%%%%%%%%%%%%%%%%%%%%%%%%%%%%%%%%%%%%%%
%% The following lines show an example how to produce a bibliography

\bibliography{aji}

\end{document}